\documentclass[review]{elsarticle}

\usepackage{hyperref}

\journal{Journal of Photonics and Nanostructures}

\begin{document}

\begin{frontmatter}

\title{High-transmissivity Silicon Visible-wavelength Metasurface Designs based
on Truncated-cone Nanoantennae}

\author{Krupali D. Donda, Ravi S. Hegde}
\address{Department of Electrical Engineering, \\
Indian Institute of Technology, Gandhinagar, 
Gujarat, India,  382355 \\
\url{hegder@iitgn.ac.in}}


\begin{abstract}
High-transmissivity all-dielectric metasurfaces have recently attracted attention towards
the realization of ultra-compact optical devices and systems. Silicon based
metasurfaces, in particular, are highly promising considering the possibility of monolithic
integration with VLSI circuits.  Realization of
silicon based metasurfaces operational in the visible wavelengths remains a
challenge.  A numerical study of silicon metasurfaces based on stepped truncated
cone shaped nanoantenna elements is presented. Metasurfaces based on the stepped
conical geometry can be designed for operation in the 700nm to
800nm wavelength window and achieve full cycle phase response (0
to pi with an improved transmittance in comparison with previously reported
cylindrical geometry~\cite{Yu2015a}. A systematic parameter study of the
influence of various geometrical parameters on the achievable amplitude and
phase coverage is reported.  
\end{abstract}


\end{frontmatter}

\section{Introduction}

 The
 metasurface~\cite{Genevet2015,Kildishev2012,Lin2014,Meinzer2014,Minovich2015},
 a paradigm-shifting concept, promises the next-generation of flat, integrable
 devices for the manipulation of optical wavefronts. It is a spatially
 heterogeneous array of nanoscale resonant elements (called meta-atoms) that can
 in general alter the amplitude, phase, spectrum and polarization values of an
 incident wavefront in a short propagation distance and with a sub-wavelength
 scale in the transverse plane~\cite{Minovich2015}.  The metasurface
 concept~\cite{Yu2011a,Yu2014b} has been primarily explored in connection with
 plasmonic nanoantenna in the recent past~\cite{Kildishev2012}.  Traditional
 plasmonic materials, however,  are not compatible with the CMOS process flow,
 and,  additionally, the high absorption~\cite{Khurgin2015} is problematic in
 transmission based applications~\cite{Aieta2012}.   Recently it was suggested
 to replace the plasmonic metals used for nanoantenna with high-index
 dielectrics resulting in the so called all-dielectric
 nanoantenna~\cite{Krasnok2014}. These nanoantenna do not strongly enhance the
 local electromagnetic field as do plasmonic nanoantenna, but they suffer far
 lower amounts of dissipative loss and offer magnetic response at higher
 frequencies~\cite{Miroshnichenko2011,Krasnok2015,Krasnok2014a,Krasnok2013}.
 Dielectric nanoantenna have been traditionally studied in the microwave
 community as dielectric resonator antenna~\cite{K.M.Luk;2003}.  

By using
 silicon based nanoantenna, reflectionless phase control of incident wavefront
 (Huygens' surface) in a transmissive metasurface was reported recently and
 offered a much higher efficiency in comparison to a plasmonic
 metasurface~\cite{Decker2014, Decker2014a,Staude2013}. High index dielectric
 nanoantenna arrays can also be considered as two dimensional high-index
 contrast subwavelength diffraction gratings; various optical wavefront
 manipulation possibilities have been demonstrated with these so called HCTA
 (High Contrast
 Transmitarrays)~\cite{Arbabi2014b,Arbabi2014d,Arbabi2016,Arbabi2015}.  Most of
 these reports have been demonstrated at near IR operating wavelengths where
 they have achieved high transmission efficiencies. Scaling the operational
 wavelength of silicon metasurfaces into the visible region is attractive as
 they can be combined with silicon photodetectors and elctronic postprocessing. 
 Currently reported silicon metasurfaces using cylindrical nanoantenna as
 meta-atoms~\cite{Yu2015a,Khorasaninejad2016} have seen worse performance in the
 visible region as compared to near IR region.  It has been speculated that the
 higher material absorption losses of silicon and increased mutual coupling
 could limit its performance
 \cite{Yu2015a,Khorasaninejad2016} prompting researchers to look into other
 material systems like Titanium dioxide~\cite{Yu2015a,Khorasaninejad2016}. 

While alternate semiconducting high-index materials are one pathway towards
realization of all-dielectric visible wavelength metasurfaces, silicon remains 
an attractive option particularly for chip-scale integration of metasurfaces.
We propose here a bilayered stepped truncated conical geometry as the meta-atom
and compare its performance with a simple cylindrical shape using silicon as the
material. We show that the proposed design performs better than the cylindrical
geometry in the visible wavelength window of 700nm to
800nm; however, this comes at the cost of added fabrication
complexity. The truncated cone shape has been previously explored in
connection with RF dielectric resonator antenna~\cite{Kishk2002} and has been
found to have several favorable properties including a wider operational
bandwidth. The paper begins with a description and exploration of the spectral
response of the proposed stepped cone nanoantenna in section~\ref{sec:struct}.
In section~\ref{sec:results}, the amplitude and phase response of this structure
and the dependence on various geometrical parameters is described before
concluding the paper in section~\ref{sec:conclusion}.

\begin{figure}[htbp] 
  \centering 
  \includegraphics[width=\textwidth]{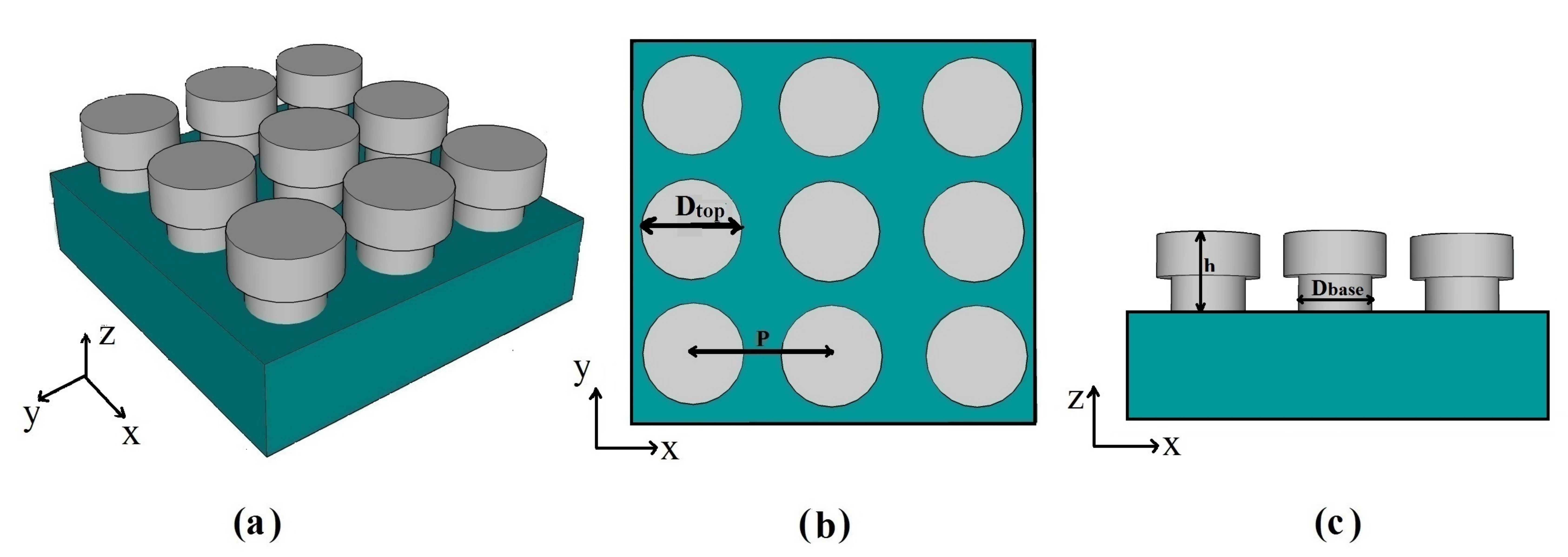}
  \caption{Schematic of the proposed inverted cone bilayer dielectric
    metasurface consisting of Silicon nanoantenna of total height $h$ in air on
    a fused silica substrate with the various geometrical parameters labelled. 
  (a), (b) and (c) show the perspective, top and side views
respectively.} 
\label{fig:structure} 
\end{figure}

\section{The stepped-conical silicon nanoantenna}\label{sec:struct}

\subsection{Structure Description}

The proposed bilayer dielectric metasurface consists of an array of
heterogeneously sized nanoantennae arranged in a regular rectangular lattice.
The schematic shown in figure~\ref{fig:structure} can be thought of as a section
of this spatially variant metasurface. The individual nanoantenna is shaped in
the form of a truncated two stepped cone and is characterized by four
geometrical parameters: the diameter of the top cone $D_{top} = 2R_{top}$,  the
diameter of the bottom cone $D_{bottom} = 2R_{bottom}$ and, the respective
heights $h_{top}$ and $h_{bottom}$.  Throughout this paper, we consider a
uniform rectangular lattice of periodicity $P$ and heights $h_{top}$ and
$h_{bottom}$; the diameters  $D_{top}$ and $D_{bottom}$ are, however, free to
change throughout the extent of the metasurface.  Furthermore, we also restrict
our attention to the case where the thickness of the metasurface or the total
height $h$ (equal to $h_{top} + h_{bottom}$) is 130nm.  The
height of 130nm was chosen so that the primary resonances occur
in the red spectral range of 700nm to 800nm
where intrinsic absorption loss of amorphous silicon is small~\cite{Yu2015a}. 

The numerical simulations were performed with the Finite Element Method (FEM)
based frequency solver in the commercial software suite CST Microwave studio.
The extinction cross section calculation uses open boundaries on all sides with
a plane wave illumination. The extinction cross sections are calculated by
summing the scattering and the absorption cross sections. The scattering cross
section is obtained by integrating the total scattered power by first generating
the far fields using far field transformations applied to the near fields
captured at the open boundaries.  On the other hand, the reflectance and
transmittance spectra for the  array are obtained by using periodic boundary
conditions and Floquet port excitation.  Tabulated values reported in the
literature were used to model the dispersive permittivity of
silicon~\cite{PALIK1985} and a constant refractive index of 1.45 is assumed for
the silica substrate. The solver uses a tetrahedral meshing scheme, starts out
with a step size of 4 per smallest simulation wavelength and a minimum of 10 for
any given edge. The solver iteratively refines the mesh and converges to a final
mesh geometry. 

\begin{figure}[htbp] 
  \centering
  \includegraphics[width=\textwidth]{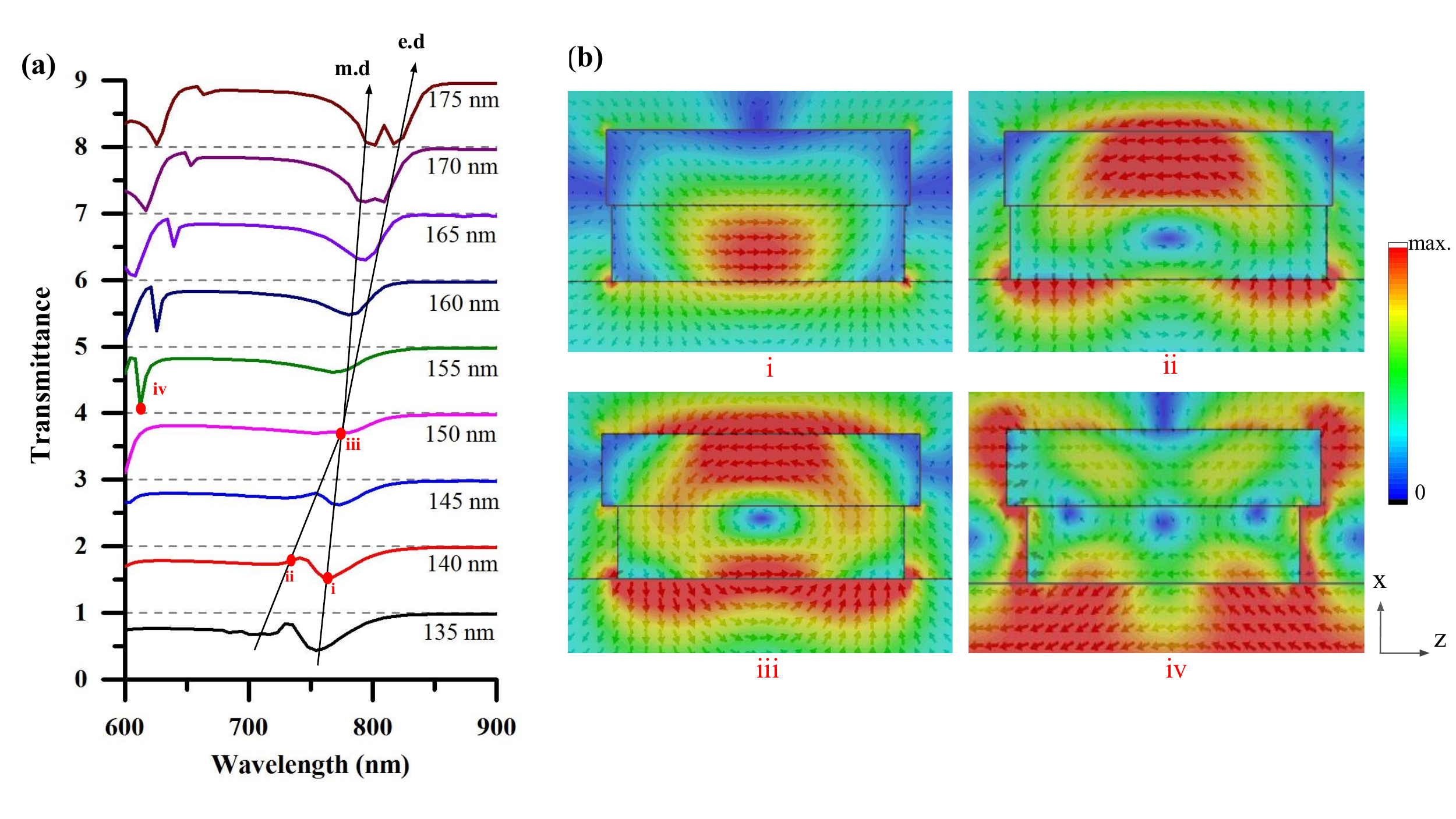} 
  \caption{(a) Transmission spectra for the stepped conical dielectric
    nanoantenna geometry. The structure is excited with a normally incident
    plane wave illumination from the top (top radius is varying fromhttps://www.sharelatex.com/project/58ec81482836187d713d78d5
    145nm to 175nm and the base radius
    135nm is held constant).  The total height $h$ is
    130nm and periodicity $P$ is $D_{top}+115$.  Guide lines are
    drawn that track the magnetic dipolar (md) and electric dipolar (ed)
    resonance wavelengths. (b) At 4 selected resonance wavelengths (marked (i),
  (ii), (iii) and (iv) in  (a)), the near field plots of the electric field is
shown for a $x-z$ cut of the structure.}
\label{fig:fieldplots}
\end{figure}

\subsection{Optical response of stepped cone} 

The spectral response of a nanoscale resonator made of a high-index dielectric
material is dominated by the magnetic and electric dipole
resonances~\cite{Alexandr} that are induced in it by the incident field. This is
seen most easily in the resonances of a spherical particle made of silicon (see
figure S1). A formal mathematical treatment of the electromagnetic scattering
behaviour of a spherical particle (reported by G. Mie in 1908~\cite{Mie1908})
predicts the existence of magnetic dipole resonances~\cite{Alexandr}.  In
particular, for a spherical particle made of a high-refractive index material,
the magnetic dipole frequency is lower in comparison to the electric dipole
frequency~\cite{Permyakov2015}. For a spherical particle,  the magnetic dipole
resonance frequency occurs at the free space wavelength $\lambda$ when
$\lambda/n(\lambda)$, the wavelength in the particle equals the diameter of the
particle $2 R_s$ (here $n(\lambda)$ is the wavelength dependent refractive index
of the scattering medium).  The orientation of the electric field becomes
exactly anti-parallel at the sphere boundaries at this condition resulting in a
circulating induced current in it; the circulating current results in the
magnetic dipole radiation and hence an increased scattering at this particular
frequency.  At the electric dipole resonance condition, the electric field lines
exhibit behavior similiar to that of an electric dipole (see figure S1).  

A nanoantenna which is non-spherical also exhibits magnetic and electric dipolar
modes.  The numerical simulation results shown in figure S2 (b and c) clearly
show the similiarity of the magnetic dipolar resonance occuring in the case of
the stepped cone structure and that of the nanosphere. In particular, the
telltale sign is the presence of the circulating electric fields.  In
figure~\label{fig:fieldplots}, various bilayerd cones are considered with
changing top radius. Widening the top radius redshifts both the magentic and the
electric dipolar resonances with the redshift of the electric dipolar resonances
being much stronger. This causes a condition whereby both the resonances occur
at the same wavelength (in this case occuring for $R_{top} =$ 150nm at the
wavelength of 780nm. It has been reported that when such an
overlap occurs, the backscattering is completely cancelled in favor of forward
scattering~\cite{Staude2013}. Thus while the transmission drops at wavelengths
where the magnetic and electric dipolar resonances individually, it is seen to
peak when the resonances overlap.

\section{Results and Discussion} \label{sec:results} 

\subsection{Amplitude and phase response}

\begin{figure}[htbp] 
  \centering 
  \includegraphics[width=\textwidth]{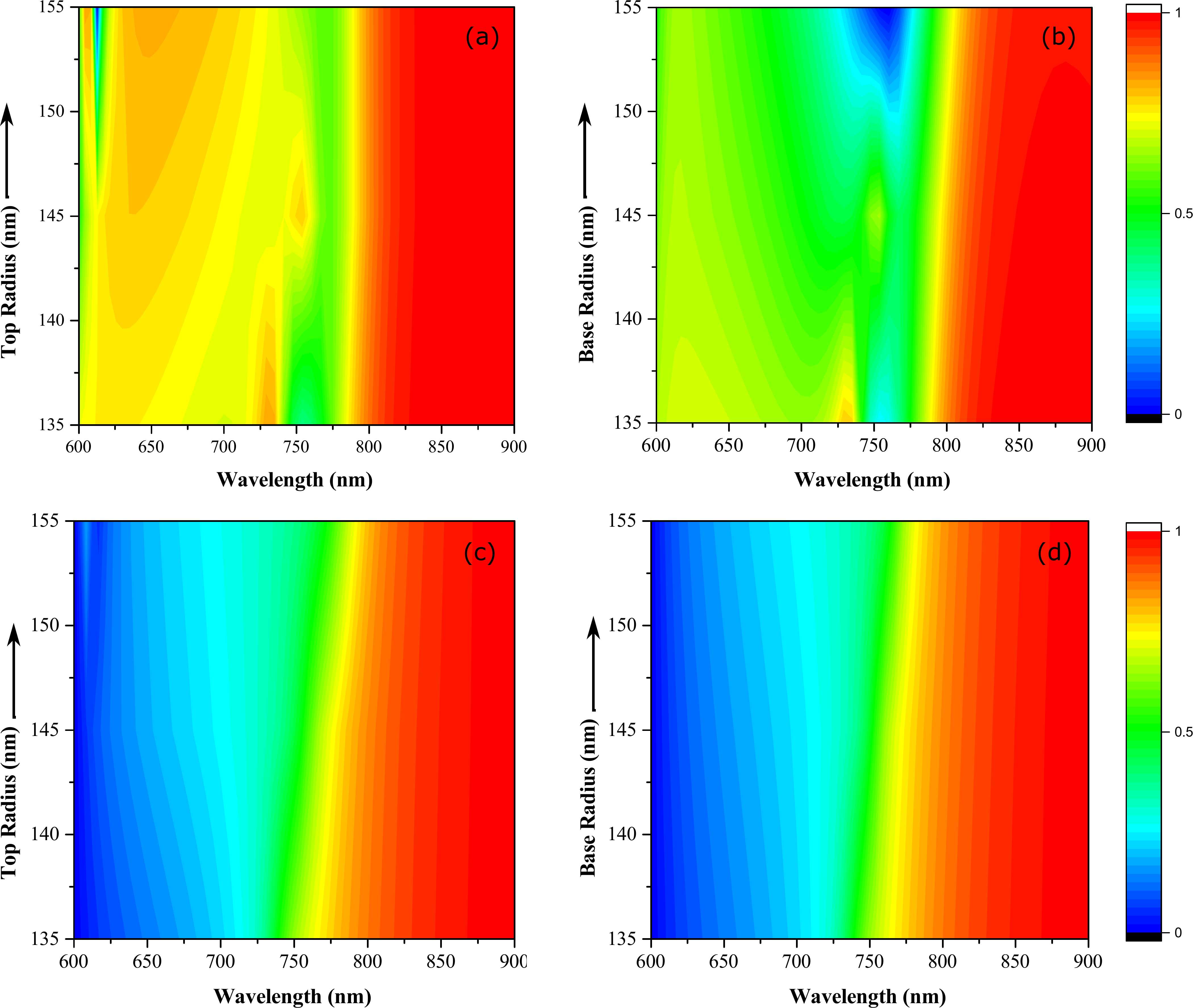} 
  \caption{ (a,c) Transmittance and phase response of inverted nanocone. Base
    radius is fixed at 135nm and top radius is changing from
    135nm to 155nm.  (b,d) Transmittance and
    phase response of straight nanocone. Top 
    radius is fixed at 135nm and base radius is changing from
    135nm to 155nm.
The total height $h$ is 130nm and periodicity $P$ is
$D_{top}+115$nm.}
\label{fig:strandinv} 
\end{figure}

Consider the transmission amplitude and phase response for the nanocone
structure. Two variations are possible: inverted and straight lying nanocones
with respect to the substrate. Figure~\ref{fig:strandinv} shows the
transmittance (the amplitude transmission is squared to obtain the
transmittance) and phase response (expressed in multiples of $2\pi$) of inverted
and straight cone structures respectively for wavelength range
600nm to 900nm.  A full range of phase coverage
is possible as noted in  (b) with high transmission efficiency (a). Here,
resonance is shifting towards lower frequencies linearly with increase in the
top diameter. The optimal transmission case is reached for top radius of 145 nm
with peak transmission $\ge$ 95\%.  In case of straight cone, it is observed
that the overall transmission is poorer than that for inverted stepped cone.
Furthermore, a full phase coverage is possible for a wider range of sizes for
the inverted cone. The inverted orientation is thus considered throughout this
article for systematic investigation.   

While the nanocylinder also achieves a full phase coverage range, the inverted
nanocone outperforms the cylinder in terms of the transmittance as seen in
figure~\ref{fig:cylcomp}.  Two sets of data are plotted: each set considers
shapes with a constant base radius but changing top radius. It is seen that the
nanocone performs better than the cylinder in each case.

\begin{figure}[htbp] 
  \centering 
  \includegraphics[width=\textwidth]{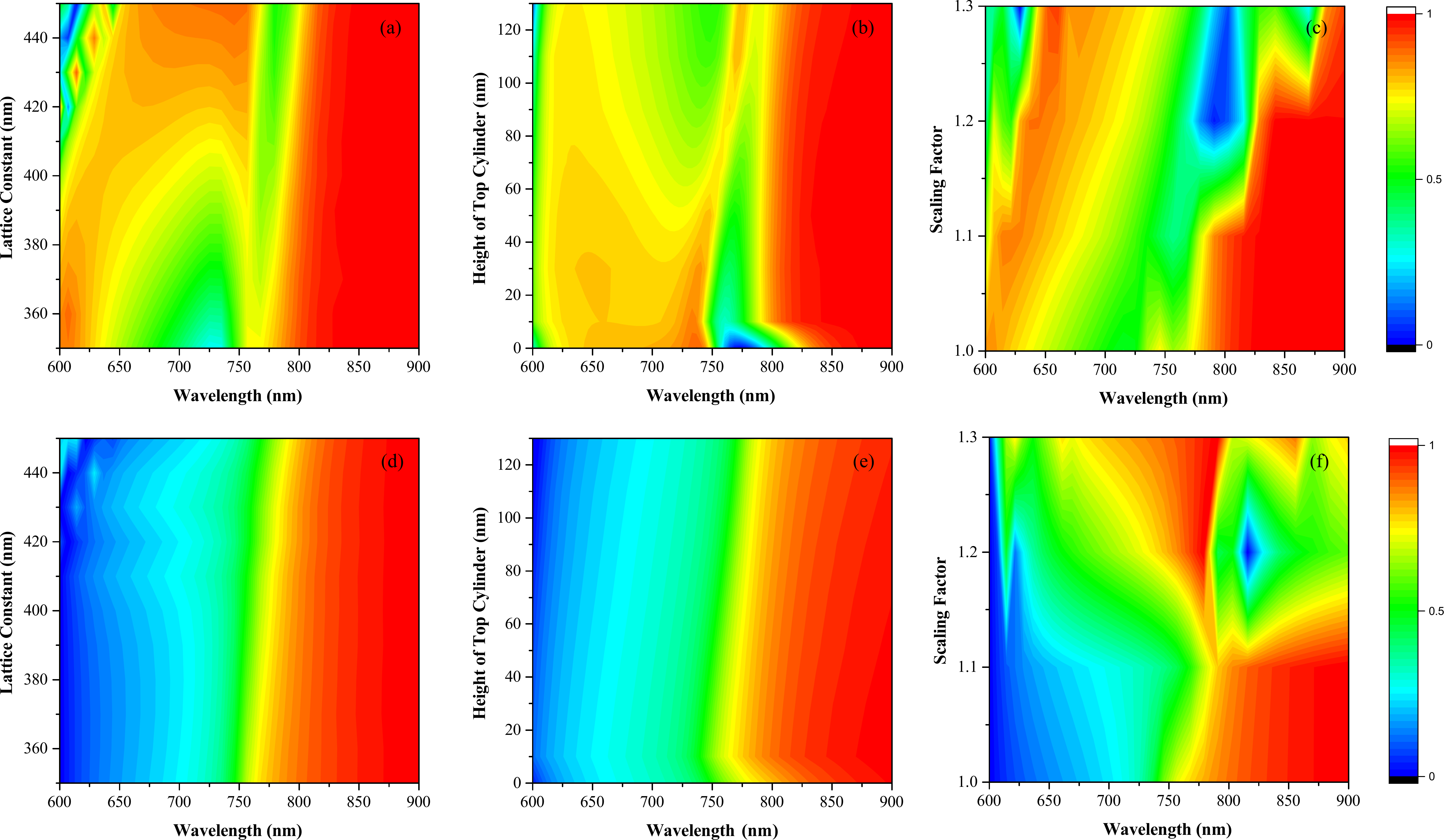} 
\caption{ (a,d) Transmittance and phase response of inverted nanocone with base radius
  135nm and top radius 145nm for variation in
  the inter-antenna spacing. The heights $h_{top}$ and $h_{bottom}$ are equal
  and the total height $h$ is 130nm. (b,e) Transmittance and
  phase response of inverted nanocone with base and top radius and total height
  same as in (a, d) but with varying $h_{top}$.  (c,f) Transmittance and phase response of inverted nanocone 
  when all the geometric properties $P, D_{top}, D_{bottom}, h_{top}$ and
  $h_{bottom}$ are scaled uniformly. The scaling factor 1 denotes the case of  
  base radius 135nm and top radius 145nm, $h$ is 130nm and periodicity 405nm. }
\label{fig:parsweep} 
\end{figure}

\subsection{Influence of geometric parameters}

This section discusses the influence of various geometrical parameters on the
response of the inverted nanocone; specifically, the influence of inter-antenna
spacing, height ratios and uniform geometrical scaling.  As seen in
figure~\label{fig:parsweep}(a), as the periodicity increases, the main resonant
dip is slightly shifted toward lower frequencies from 750nm to
800nm and a narrow spectral feature, arising out of another
resonance is created in the wavelength region of 600nm and
650nm. Periodicity values of 400nm to
450nm are seen to provide the highest overall transmittance.
Phase coverage, however,  is seen to be unaffected with periodicity change as
seen in figure~\label{fig:parsweep}(b).  As seen in
figure~\label{fig:parsweep}(b), a height ratio of 1:1 is seen to provide the
best overall transmittance. Again, the phase coverage is not strongly influenced
by this factor as seen in figure~\label{fig:parsweep}(e). Uniform scaling of the
nanocone is expected to shift the operating wavelength to a higher wavelength.
However, unexpectedly, this expectation is not borne out by the results seen in
figure~\label{fig:parsweep}(c) and (f).  

\subsection{Metasurface design at a specific wavelength}

\begin{figure}[htbp] 
  \centering 
  \includegraphics[width=\textwidth]{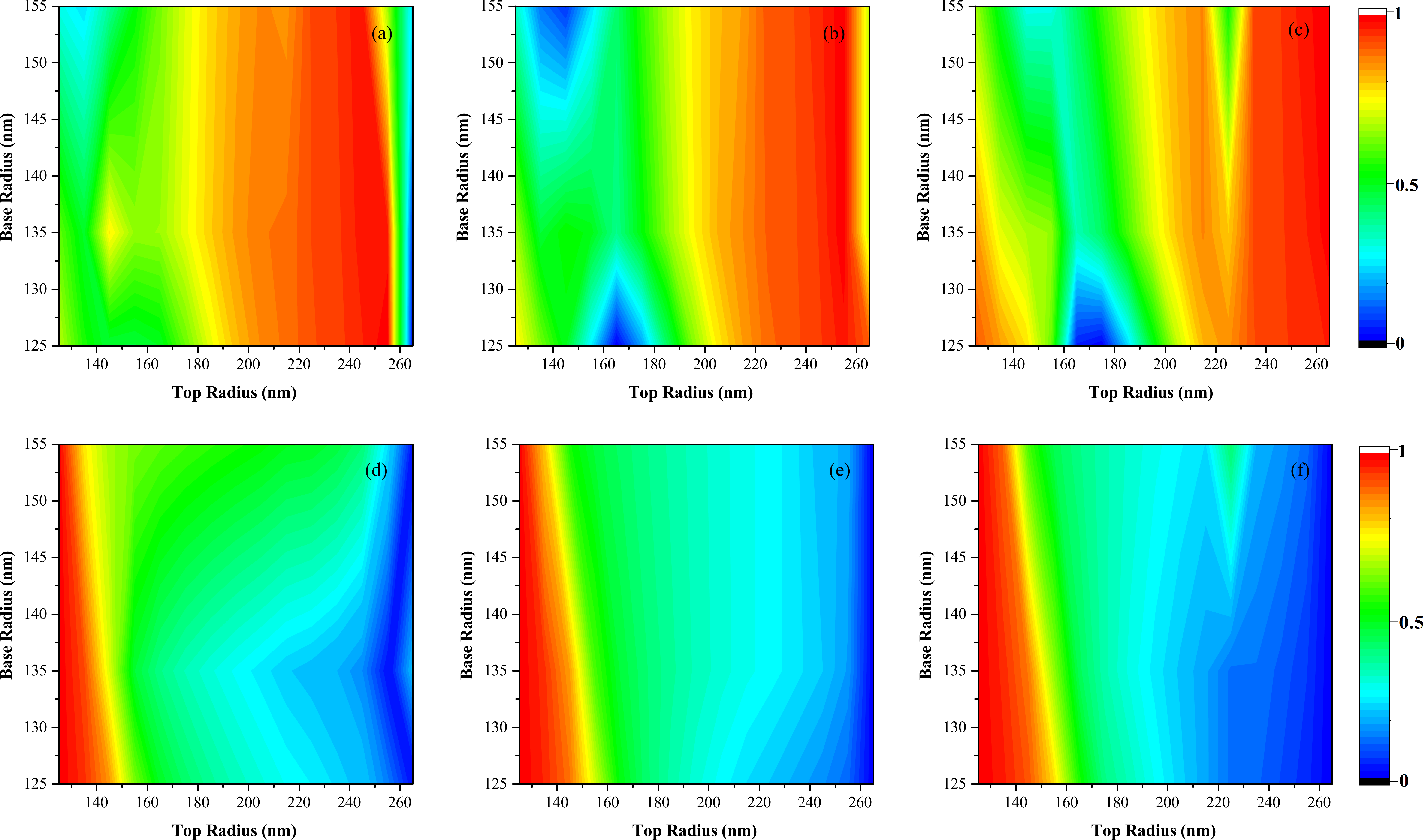} 
\caption{Transmittance and Phase response for various combinations of base
  and top radii at the wavelength of 770nm where there is the
  best overlap of the magnetic and electric dipole resonances. Comparison with
  nearby wavelengths is also shown. (b,e) transmittance and phase response at   
  770nm, (a,d) 760nm and (c,f)
  780nm. Height h is 760nm and the periodicity
  P = D + 115}
\label{fig:speclamb} 
\end{figure}

The optimal transmission case is reached for top radius of 145 nm and base
radius of 135 nm with peak
transmission $\ge$ 95\% at the wavelength of 770nm as can be
seen in figure~\ref{fig:strandinv}.  At this wavelength, a precise overlap of
the electric and magnetic dipole resonances occurs and leads to near unity
transmission and zero reflection Huygens surface~\cite{Staude2013}. 
To design a metasurface with a full phase control at this specific wavelength,
we can vary the geometrical parameters around the optimal
combination~\cite{Yu2015a}. Note that as the geometrical parameters change to
provide a particular phase, the transmittance value may drop from the peak
value. Figure~\ref{fig:speclamb} shows the transmittance and phase response for
various combinations of geometrical parameters for the inverted truncated cone. 
The overall device efficiencies reported, for
instance, in a nanocylinder based metasurface beam deflector is around
45\%~\cite{Yu2015a}. The addition of a second geometrical parameter makes it
possible to optimally pick geometrical parameters to maximize the transmittance
at any given phase angle. This will lead to improvement of overall device
efficiency. 

The transmittance and phase response changes for a shift in the operational
wavelength decides the operational bandwidth of the metasurface device. From
figure~\ref{fig:speclamb}, we see that the total operational bandwidth is indeed
quite limited.  Yu and coworkers~\cite{Yu2015a} report on a peculiar observation that
experimentally observed transmittance is better than the simulated ones. In the
simulations, it is observed that he magnetic dipole resonance is fairly broader
in comparison with the electrical case,  and, furthermore, that the electrical
resonance is less spatially confined. This decreases the overall spectral
overlap between the magnetic and dipole resonances. In experimental studies,
various non-idealities serve to broaden the electric dipole resonance. Thus, the
operational bandwidth is expected to be better than that observed in
simulations.

\section{Conclusion} \label{sec:conclusion}

A design modification is proposed for nanocylinder based silicon dielectric
metasurfaces that can improve their overall observed transmission efficiencies
for visible-wavelength operation. The efficiency improvement however requires
the fabrication of two aligned layers. We note, however, that several workers
have already reported metasurfaces that use a bilayered geometry~\cite{Tate2009,
Taubert2012, Bott2014} with requirement of precise nanoscale alignments
requirements between the two layers.  The most direcly relevant reported
structures are the so called ``mushroom''
nanopillars~\cite{YWangHHuJShao2014,Forati2016,Kang2016}. By the use of a
sacrificial layer~\cite{YWangHHuJShao2014}, the second layer can be fabricated
(see figure S3).  Fabrication of the bilayer metasurface could potentially lead
to misalignment in the two component cylinders. We have studied the effect on
transmittance and phase response due to misalignment of top cylinder with
respect to bottom cylinder (see figure S4).  The misalignment leads to a
translation of one center with respect to the other.  For small offsets of up to
25nm, there is a negligible alteration in the transmittance and
phase response. 

Future studies can consider other non-symmetrical geometries. Although, the
reported metasurface design allows a wider simultaneous control of amplitude
and phase; not all possible combinations are obtained. We are investigating
other geometries to allow fully arbitrary selection of both phase and
amplitude change. The numerical study assumes the unit-cell approach whereby
the mutual coupling between neighboring elements is assumed to be
negligible.  The mutual coupling effect is expected to become more dominant
as the wavelength is further reduced below 700nm. Thus the reported designs
may not scale below 700nm. Future studies will investigate the influence of
mutual coupling and devise design strategies that take these into
consideration. 

\section*{Acknowledgment}
The authors acknowledge support from the Department of Science and
Technology, Govt. of India through the Extramural grant SB/S3/EECE/0200/\\
2015.

\section*{REFERENCES} 
\bibliographystyle{unsrt} 
\bibliography{all.bib}
\end{document}